\documentclass[useAMS,usenatbib]{mn2e}
\usepackage{graphicx,longtable,color}
\usepackage{mathrsfs}
\usepackage{epsfig}
\usepackage{natbib}
\usepackage{color}
\usepackage{float}
\usepackage{amsmath}
\usepackage{times}
\usepackage{upgreek}
\usepackage[varg]{txfonts}
\usepackage{enumerate}
\usepackage{verbatim}
\usepackage{booktabs}
\usepackage{multirow}
\usepackage{amssymb}
\usepackage{placeins}

\title[Galaxy alignment as a probe of large-scale filaments]{Galaxy alignment as a probe of large-scale filaments}
\author[Yu Rong, Yuan Liu, Shuang-Nan Zhang]{Yu Rong$^{1,2}$\thanks{E-mail: rongyu@ihep.ac.cn}, Yuan Liu$^{1}$, Shuang-Nan Zhang$^{1,3}$\thanks{E-mail: zhangsn@ihep.ac.cn}\\
$^{1}$Key Laboratory of Particle Astrophysics, Institute of High Energy Physics, Chinese
Academy of Sciences, Beijing, China\\
$^{2}$University of Chinese Academy of Sciences, 19A Yuquan Road, 100049 Beijing, PR China\\
$^{3}$National Astronomical Observatories,
Chinese Academy Of Sciences, Beijing, China}

\begin{document}
\maketitle

\begin{abstract}
The orientations of the red galaxies in a filament are aligned with the orientation of the filament. We thus develop a location-alignment-method (LAM) of detecting filaments around clusters of galaxies, which uses both the alignments of red galaxies and their distributions in two-dimensional images. For the first time, the orientations of red galaxies are used as probes of filaments. We apply LAM to the environment of Coma cluster, and find four filaments (two filaments are located in sheets) in two selected regions, which are compared with the filaments detected with the method of \cite{Falco14}. We find that LAM can effectively detect the filaments around a cluster, even with $3\sigma$ confidence level, and clearly reveal the number and overall orientations of the detected filaments. LAM is independent of the redshifts of galaxies, and thus can be applied at relatively high redshifts and to the samples of red galaxies without the information of redshifts.
\end{abstract}
\section{Introduction}

Galaxies are not distributed randomly in the cosmic web \citep{Joeveer78,Zeldovich82,Shandarin83,Einasto84,Bond96,Aragon10}, but are arranged in filaments and sheets surrounding cosmic voids and connecting clusters of galaxies \citep{Pimbblet04,Aragon10,Jasche10,Tempel14b,Cautun14}. The properties of filaments affect the abundance, shape, and evolution of galaxies \citep{Aragon07b,Hahn07a,Hahn07,Libeskind12,Libeskind13,Cautun13,Tempel13,Tempel13b}, and depend on the properties of the initial density fluctuations generated in the very early Universe. Therefore probes of the large-scale filaments enable us to test current physical and cosmological theories.

Probes of the filaments require efficient algorithms for finding filaments. Several sophisticated algorithms for identifying filaments, both in the three dimensions and two dimensions, have been developed. Generally, there are three types of algorithms to identify a filament based on: (1) the distribution of galaxies or clusters of galaxies \citep{Peebles80,Peacock98,Novikov06,Aragon07a,Sousbie11,Shandarin12,Cautun13}; (2) the gravitational tidal tensor -- the Hessian of the gravitational potential \citep{Lee08,Forero09,Bond10a,Bond10b,Wang12}; and (3) the velocity field induced from the dynamics of the underlying density field \citep{Hahn07,Shandarin11,Hoffman12,Libeskind12,Libeskind15,Falco14}. Recently, \cite{Falco14} found that the line-of-sight velocities $v_{\rm{los}}$ of the galaxies around a cluster of galaxies with the projected distances to the cluster center satisfying $2.5r_{\rm{vir}}\lesssim r\lesssim 8r_{\rm{vir}}$, where $r_{\rm{vir}}$ is the virial radius, are as a function of $r$, if the galaxies are arranged in one filament or sheet, since these galaxies are gravitationally affected by the cluster. Therefore they plotted the $(r,v_{\rm{los}})$ map for the galaxies, and identified the filamentary structures on the map as filaments or sheets (see the paper of Falco et al. 2014 for details). They proved that the line-of-sight velocities can be probes of filaments and sheets.

In this paper, we develop a new method of identifying filaments using the orientations of galaxies. \cite{Tempel13b,Tempel15} found that the spin axes of bright spiral galaxies have a weak tendency to be aligned parallel to filaments, while the major axes of elliptical/S0 galaxies are significantly aligned with their host filaments. Therefore the galaxies alignment may be an additional probe of filaments. We describe the method, and apply it to the galaxies assembled around the Coma cluster in section 2. The results are compared with the detected standard filaments using the method of \cite{Falco14} in this section, and discussed in section 3. In section 4, we summarize the work. We adopt the WMAP7 cosmological parameters: $\Omega_{\rm{M}}=0.27$, $\Omega_{\rm{\Lambda}}=0.73$, and $H_0=71\ \rm km\ s^{-1}\ Mpc^{-1}$.

\section{Identification of Filaments by Galaxies Alignment}
\subsection{Galaxies in a filament}

For the galaxies arranged in a filament, the major axes of red galaxies and spin axes of blue galaxies are related to their host filaments, e.g., the major axes of the elliptical/S0 (red) galaxies are significantly aligned with the orientations of the filament \citep{Tempel15}. Therefore, according to the distribution of orientations of the projected red galaxies, we may identify the filaments in two-dimensional (2D) images.

\begin{figure}
\centering
\includegraphics[scale=0.4]{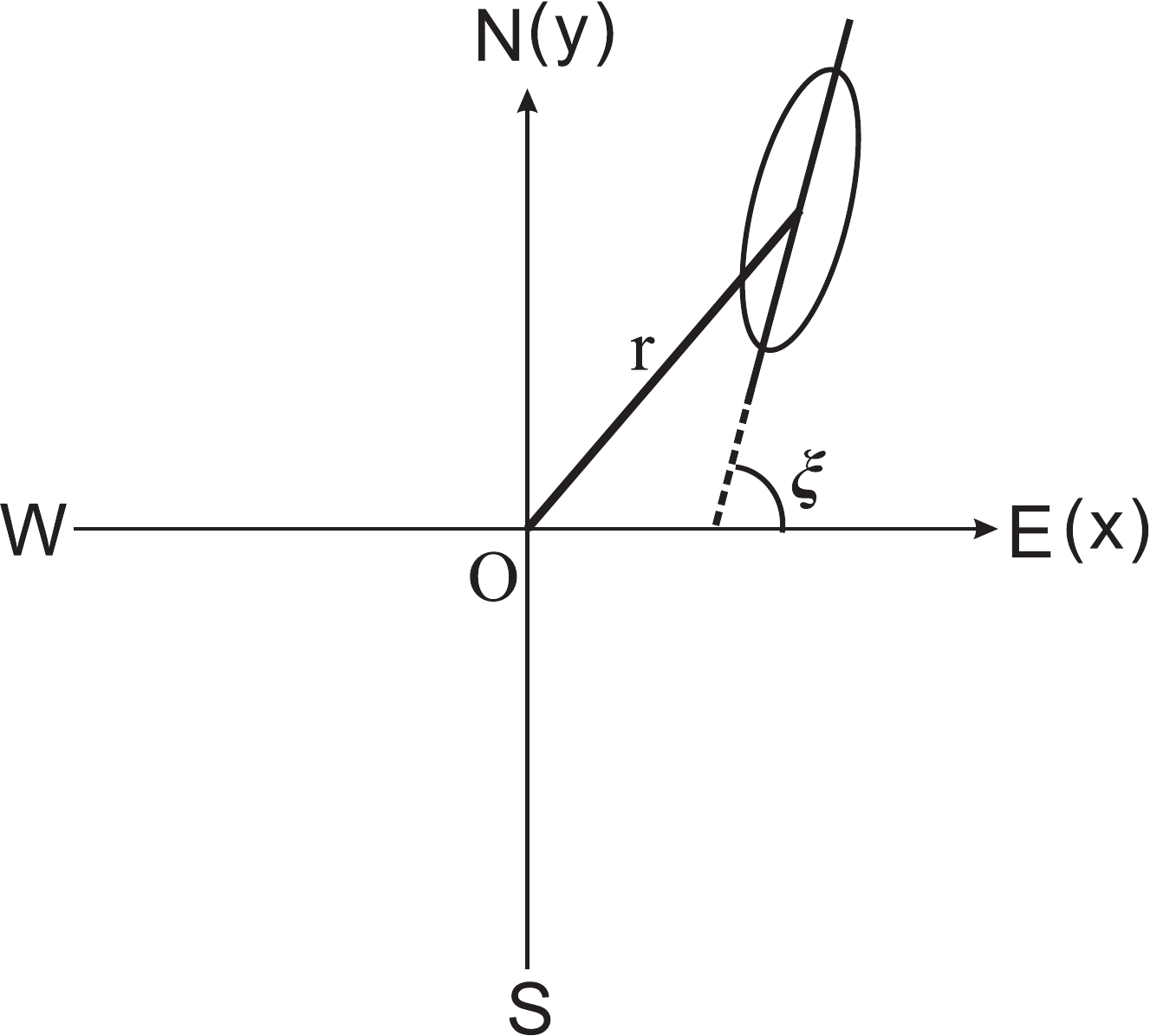}
\caption{Illustrations of the coordinate system, position angle $\xi$, and projected distance $r$. O denotes the center of a cluster of galaxies.}
\label{ske_pa}
\end{figure}

In order to quantitatively indicate the orientations of the projected ellipses, we define the east direction in the celestial coordinate system as $x$-axis, and the north direction as $y$-axis, and define the position angle of a red galaxy, $\xi$, as the angle between the major axis of the projected ellipse and the $x$-axis, as shown in Fig.~\ref{ske_pa}. The range of $\xi$ is $0\--90^{\circ}$.

Analogous to the work of \cite{Falco14} using the $(r,v_{\rm{los}})$ map, we can plot the $(r, \xi)$ map for the projected red galaxies in an image, where the horizontal coordinate $r$ denotes the distance between a projected ellipse to the origin of the coordinates, and the longitudinal coordinate is the position angle $\xi$. Since we aim to identify the large-scale filaments around a cluster, the origin of the coordinates is preferentially set as the center of a cluster. The $(r, \xi)$ map sufficiently uses the 2D information of the orientations and positions of the red galaxies. If the red galaxies are not arranged in some filament structures, $\xi$ should be uniformly random within $0\--90^{\circ}$. Conversely, if the galaxies are located in a filament, we should expect the non-random distribution of $\xi$ and inhomogeneous distribution of $r$. This manner of detecting filaments is called location-alignment-method (LAM).

\subsection{Data Analysis}

We will apply LAM to the data of the galaxies around the Coma cluster, and then compare the results with the filaments obtained by the method of \cite{Falco14}. We take the galaxy NGC~4874 as the center of the Coma cluster \citep{Kent82} and origin of the coordinates, which is located at RA = $12^{\rm{h}}59^{\rm{m}}35^{\rm{s}}.7$, and DEC = $+27^{\circ}57'33''$. The galaxies used here are selected from the Sloan Digital Sky Survey data release 12 (SDSS DR12; Alam et al. 2009) with $r$-band model magnitudes (modelMag), $m_r$, satisfying $12\ {\rm{mag}}<m_r\leq 18\ \rm{mag}$, classified as galaxies photometrically by the SDSS data reduction pipelines (i.e.,photoObj.type=3), and distributed within $9^{\circ}$ from the position of the Coma center and with redshifts between 0.01 and 0.037 (i.e., velocities along the line of sight are between 3000 and 11000~km~$\rm{s}^{-1}$; Falco et al. 2014). 2590 galaxies around the Coma cluster are obtained.

\begin{figure}
\centering
\includegraphics[scale=0.49]{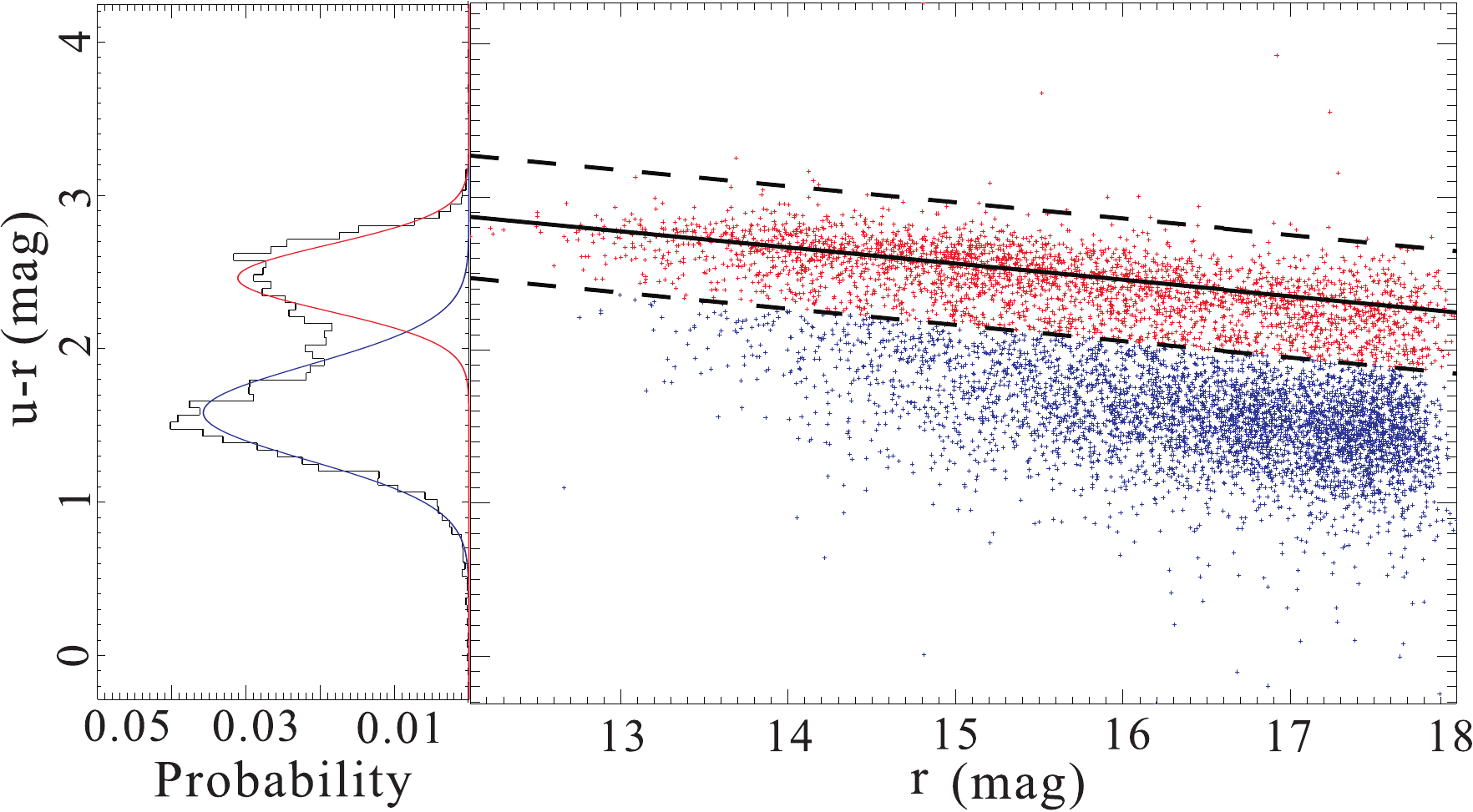}
\caption{The distribution of color and CMD of the galaxies around the Coma cluster. $u-r$ color is plotted against the $r$ magnitude of the galaxies. The elliptical/S0 galaxies and spiral galaxies are denoted by red and blue points, respectively. The black solid line denotes the ridgeline of red galaxies, and the two dashed lines confine the galaxies within $2\sigma$ color range around the ridgeline.}
\label{cmd}
\end{figure}

In order to obtain the position angles of the red galaxies, first we plot the $u-r$ versus $r$ color-magnitude diagram (CMD) and distribution of colors, $u-r$, of the 2590 numbers of galaxies, which are shown in Fig.~\ref{cmd}. Second, we select the red galaxies. The red sequence \citep{Bower92} is formed by old elliptical galaxies whose spectra show similar $4000\ \AA$ breaks (a characteristic of old stellar populations ubiquitous in elliptical galaxies; Pereira \& Kuhn 2005) resulting from photospheric absorptions of heavy elements, and the $u$ and $r$ filters nicely probe the spectral region across the $4000\AA$ break. A linear fitting to the red sequence,
\begin{equation}
(u-r)=k\cdot r+d,
\label{lf}
\end{equation}
where $k\simeq -0.106$, $d\simeq 4.154$ are the slope and intercept respectively, is applied to obtain the slope of the ridgeline \citep{Bower92} in the CMD. As shown in the distribution of colors of the galaxies, the colors of the red and blue galaxies approximately follow two Gaussian distributions, and the standard error $\sigma\simeq 0.2\ \rm{mag}$ for the colors of the red galaxies is obtained, therefore we select the galaxies confined within $2\sigma\simeq 0.4\ \rm{mag}$ around the ridgeline as the red galaxies, which are known as the collections of elliptical/S0 galaxies and dwarf ``ellipticals'' \citep{Hung10}. The rest bluer galaxies are mostly spiral galaxies. The selection criterion satisfies the Butcher-Oemler condition, i.e., blue galaxies are those at least 0.2 mag bluer than the cluster ridgeline \citep{Butcher84}. Subsequently, $\xi$ of the red galaxies are calculated by the parameters ``photoObj.deVPhi\_r'' (parameter of the de Vaucouleurs surface brightness fitting model) from the SDSS data reduction pipelines, since the de Vaucouleurs $R^{1/4}$ surface brightness profile fits well to many elliptical galaxies \citep{Vaucouleurs48}.

\begin{figure*}
\centering
\includegraphics[scale=0.6]{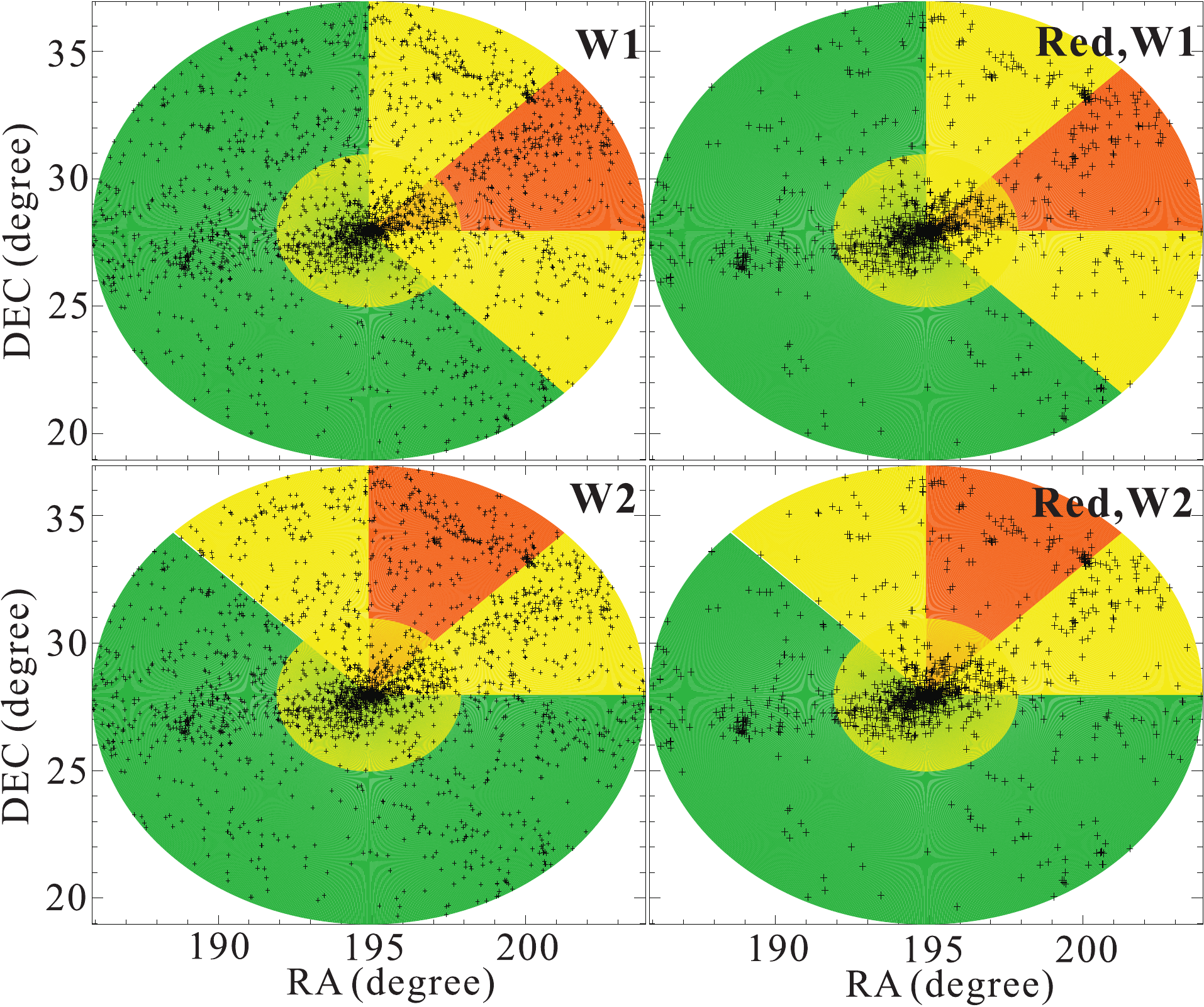}
\caption{The Coma cluster and its environment up to $9^{\circ}$ from the cluster center. The left and right panels show all the galaxies (red and blue) and only the red galaxies in this region, respectively. The region within $3^{\circ}$ from the cluster center is denoted by the central yellow round. The two selected wedges (W1 and W2), corresponding background wedges, and two wedges adjacent to each selected wedge, are colored orange, green, and yellow, respectively.}
\label{pa_filament_detect}
\end{figure*}

Our goal is to find structures confined in a relatively small region in the $x-y$ plane. Fig.~\ref{pa_filament_detect} shows the Coma cluster and its environment up to $9^{\circ}$ from the cluster center. The 536 galaxies distributed within about $1.3^{\circ}$ (corresponding to the virial radius $r_{\rm{vir}}\simeq2.3\ \rm{Mpc}$; Keshet et al. 2014) are mainly the members of Coma cluster and should be removed. Additionally, since the method of \cite{Falco14} can only detect the galaxies in filaments with $r\gtrsim 2.5r_{\rm{vir}}$, therefore the galaxies distributed within $r\lesssim 3^{\circ}$ are also removed in order to compare our results with those obtained by the method of \cite{Falco14}. All of the removed galaxies are denoted by the points in the central yellow round region in Fig.~\ref{pa_filament_detect}. The image of the galaxies except the central removed galaxies is split into eight wedges, and our goal is to detect the filaments in two wedges (W1 and W2), which are denoted by the points of the orange regions in Fig.~\ref{pa_filament_detect}. For each selected wedge, the $(r, \xi)$ map of the red galaxies in this wedge is obtained; the $(r, \xi)$ map is divided into $9\times9$ cells with $\Delta r\times \Delta \xi=1^{\circ}\times 10^{\circ}$, where $\Delta r$ and $\Delta \xi$ are the bins of $r$ and $\xi$, respectively. We need to compare the galaxy number density $n_i$ in each cell $i$, with the expected number of background galaxies in this cell $n^{\rm{bg}}_i$, to determine whether there is an excess in the $(r, \xi)$ map. The five background wedges are represented by the green regions shown in Fig.~\ref{pa_filament_detect}, and we artificially set $\xi$ of the background red galaxies uniformly random in $0\--90^{\circ}$ instead of their real orientations, since the background galaxies may be also arranged in filaments and thus grouped into some specific $\xi$ ranges. In other words, we only use the information of the background density, but not their orientations. We exclude the two wedges (yellow wedges) adjacent to each selected wedge, since any structure in the selected wedge might stretch to the closest two wedges.

\begin{figure*}
\centering
\includegraphics[scale=0.6]{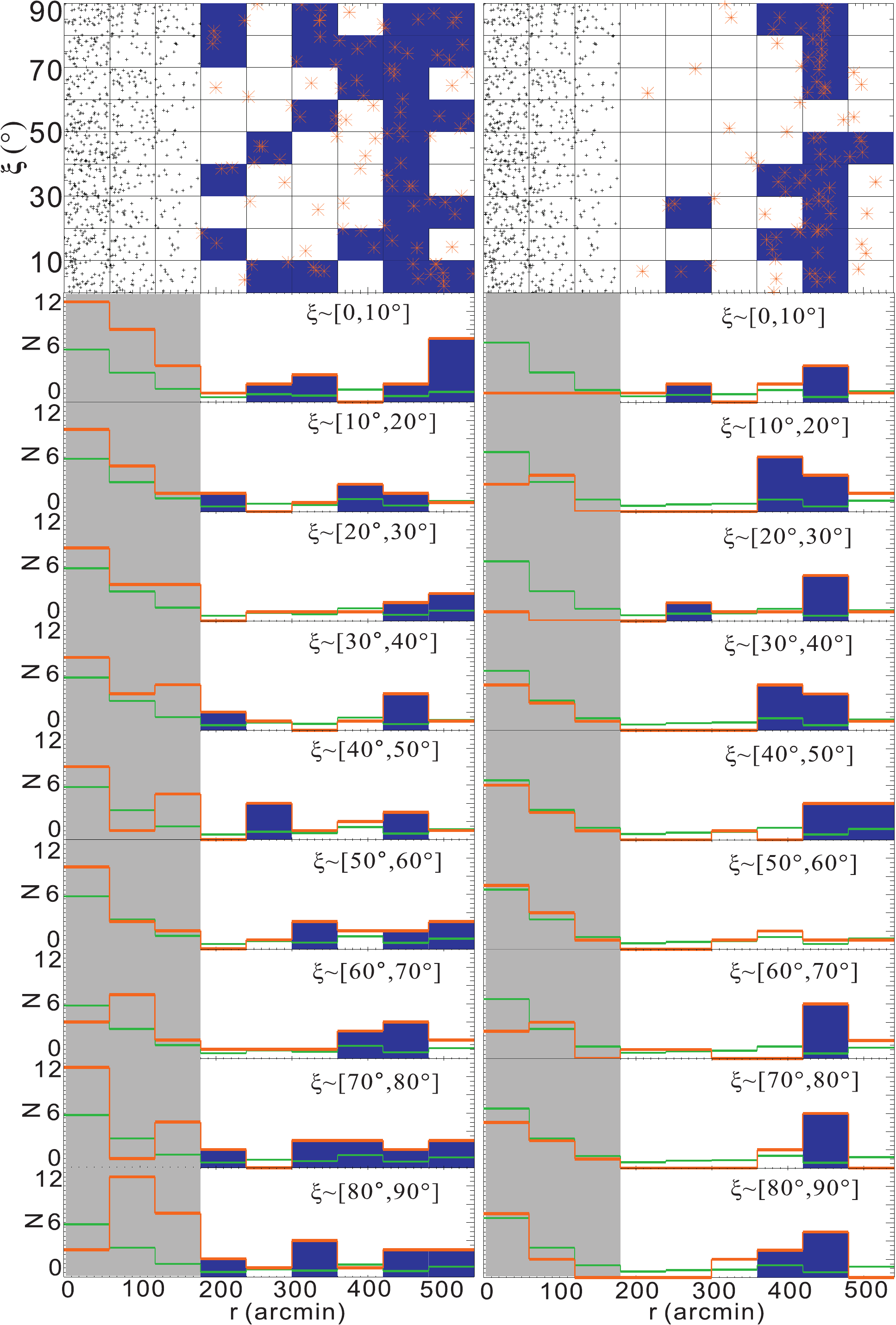}
\caption{The left and right panels show the $(r,\xi)$ map and distribution of $r$ of the red galaxies for W1 and W2, respectively. The top panel shows the $(r,\xi)$ maps for the red galaxies in the two selected wedges. The red galaxies inside and outside the central $3.0^{\circ}$ region are denoted by the black and orange points, respectively. The selected cells with the confidence level of $3\sigma$ are colored blue. The lower nine panels show the distributions of $r$ of the red galaxies within the nine $\xi$ bins. The $3.0^{\circ}$ region from the cluster center is colored gray. The orange and green histograms denote the distributions of $r$ of the red galaxies in the selected and background wedges, respectively. The final chosen cells are also denoted by the blue bins.}
\label{pas}
\end{figure*}

The excess in cell $i$ is defined as
\begin{equation}
m_i=\frac{n_i-n_i^{\rm{bg}}/5}{n_i^{\rm{bg}}/5}.
\label{ov_eq}
\end{equation}
If $m_i>0$, it means that there is an excess in the cell $i$. In order to select statistically significant excesses, we repeatedly set $\xi$ of the background galaxies $10^5$ times. Each time, the cells with $m_i>0$ are selected. During the $10^5$ times of tests, some cells are always selected, but some other cells are selected only a few times. Finally, we only choose the cells with the accumulated selected times larger than a given criteria, for instance $1\sigma$ (about 68264 times), $2\sigma$ (about 95456 times), and $3\sigma$ (about 99725 times); if a cell is chosen, all of the red galaxies in the cell are selected. Representatively, we plot the $(r,\xi)$ maps and the final chosen-cells with the confidence level of $3\sigma$ for the two selected wedges, as shown in Fig.~\ref{pas}. For each $\xi$ bin (bin of $10^{\circ}$), we also plot the histogram of $r$ of the red galaxies in this $\xi$ bin, i.e., the number of red galaxies as a function of $r$ (as denoted by the orange histogram in Fig.~\ref{pas}), and the histogram of the simulated background galaxies in this $\xi$ bin, i.e., $\bar{n}^{\rm{bg}}/5$ as a function of $r$ (as denoted by the green histogram in Fig.~\ref{pas}), where $\bar{n}^{\rm{bg}}_i$ is the mean value of $n^{\rm{bg}}_i$ during the $10^5$ times of tests. The final chosen-cells are denoted by the blue bins.

\subsection{Results and Comparison}

As described in introduction, \cite{Falco14} developed a method of identifying large-scale filaments and sheets around a cluster using the $v_{\rm{los}}$ and 2D positions of galaxies $r$, which has been proved to be robust for both the simulations and observations \citep{Falco14,Lee15}. In this paper, we use this method to identify the filaments in the same two wedges around the Coma cluster, and treat the detected filaments as the ``standard'' ones to be compared with the results by LAM. The detailed process of detecting the standard filaments can be found in \cite{Falco14}.

\begin{figure*}
\centering
\includegraphics[scale=0.6]{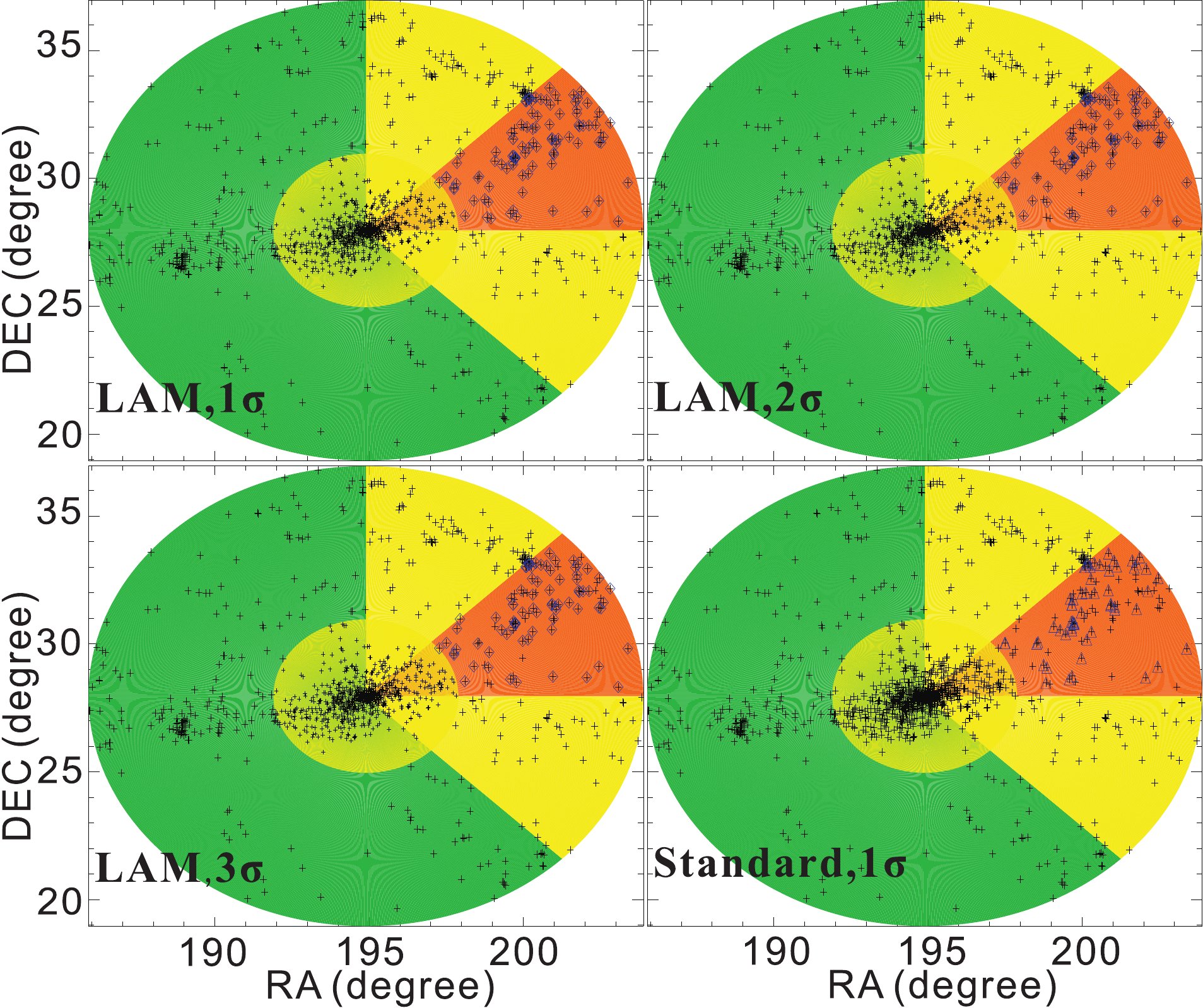}
\caption{The selected red galaxies by LAM with the $1\sigma$, $2\sigma$, and $3\sigma$ confidence levels for W1 are denoted by the blue diamonds in the upper-left, upper-right, and lower-left panels, respectively. The detected standard filaments by the method of Falco et al. (2014) are denoted by the blue triangles in the lower-right panel.}
\label{w1}
\end{figure*}

\begin{figure*}
\centering
\includegraphics[scale=0.6]{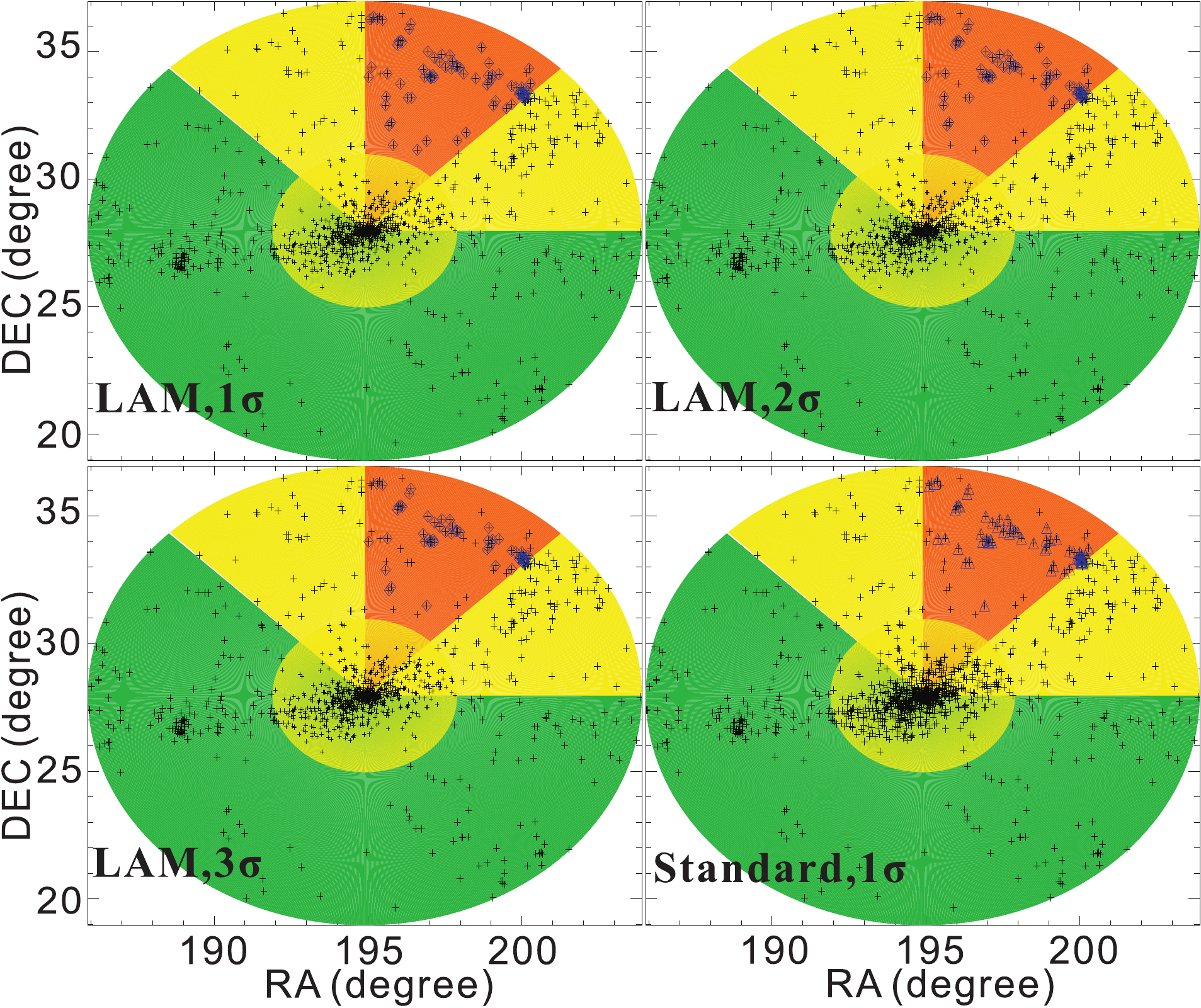}
\caption{Analogous to Fig.~\ref{w1}, this figure shows the selected red galaxies by LAM and method of Falco et al. (2014) for W2.}
\label{w2}
\end{figure*}

The red galaxies in W1 and W2 detected by LAM are shown in Figs.~\ref{w1} and \ref{w2}, and denoted by the blue diamonds. The upper-left, upper-right, lower-left panels of the two figures show the results with the criteria $1\sigma$, $2\sigma$ and $3\sigma$, respectively; and the lower-right panels show the red galaxies of the standard filaments (denoted by the blue triangles) in the corresponding wedges using the method of \cite{Falco14}. It is worth noting that the standard filaments can only be detected with the $1\sigma$ confidence level.

In Table~\ref{N_comp}, the filaments obtained by LAM are compared with the standard filaments in the two wedges. For each wedge, we count the numbers of galaxies in the detected filaments ($N_{\rm{det}}$) by LAM and standard filaments ($N_{\rm{std}}$), respectively, and the number of galaxies in both the detected and standard filaments ($N_{\rm{c}}$). Thus the fraction of the duplicated galaxies in the detected filaments, $N_{\rm{c}}/N_{\rm{det}}$, and standard filaments, $N_{\rm{c}}/N_{\rm{std}}$, are obtained. The latter suggests the detection efficiency of LAM: a higher $N_{\rm{c}}/N_{\rm{std}}$ indicates more effective LAM.

\begin{table} \scriptsize
\begin{tabular}{@{}cccc|c|c@{}}
\hline
\hline
\multicolumn{6}{c}{W1}\\
\hline
  & \multicolumn{3}{c}{LAM} & Alignment Only & No Redshifts\\
  & $1\sigma$ & $2\sigma$ & $3\sigma$ & $1\sigma$ & $1\sigma$\\
 \hline
 $N_{\rm{std}}$ & 59 & 59 & 59 & 59 & 59\\
 $N_{\rm{det}}$ & 96 & 81 & 79 & 97 & 87\\
 $N_{\rm{c}}$   & 56 & 45 & 44 & 41 & 49\\
 $N_{\rm{c}}/N_{\rm{std}}$ & 95.0\% & 76.3\% & 74.6\% & 69.5\% & 83.0\%\\
 $N_{\rm{c}}/N_{\rm{det}}$ & 58.3\% & 55.6\% & 55.7\% & 42.3\% & 56.3\%\\
\hline
\hline
\multicolumn{6}{c}{W2}\\
\hline
 & \multicolumn{3}{c}{LAM} & Alignment Only & No Redshifts\\
 & $1\sigma$ & $2\sigma$ & $3\sigma$ & $1\sigma$ & $1\sigma$\\
\hline
$N_{\rm{std}}$ & 66 & 66 & 66 & 66 & 66\\
$N_{\rm{det}}$ & 79 & 66 & 60 & 61 & 69\\
$N_{\rm{c}}$   & 63 & 55 & 54 & 42 & 58\\
$N_{\rm{c}}/N_{\rm{std}}$ & 95.4\% & 83.3\% & 81.8\% & 63.6\% & 87.9\%\\
$N_{\rm{c}}/N_{\rm{det}}$ & 79.7\% & 83.3\% & 90.0\% & 68.8\% & 84.0\%\\
\hline
\hline
\end{tabular}
\caption{The results of the detected galaxies for W1 and W2. The second to fourth columns list the results by LAM with the $1\sigma$, $3\sigma$, and $5\sigma$ confidence levels, respectively. The fifth column lists the results with the $1\sigma$ confidence level, when we only use the information of galaxies alignments. The last column lists the results without restricting the redshifts of galaxies.}
\label{N_comp}
\end{table}

According to Table~\ref{N_comp}, two points are concluded. First, with $1\sigma$ confidence level, the detection efficiency of LAM (denoted by $N_{\rm{c}}/N_{\rm{std}}$) is better than $95\%$, indicating that LAM effectively find out most galaxies found by the method of \cite{Falco14} in the filaments also with $1\sigma$ confidence level. The detection efficiency of LAM decreases with the increasing confidence level, but is always better than $75\%$, suggesting that LAM is still valid with such high confidence levels. Second, $N_{\rm{c}}/N_{\rm{det}}$ is relatively low, suggesting that LAM detects some other galaxies in filaments which cannot be found by the method of \cite{Falco14}.

\begin{figure*}
\centering
\includegraphics[scale=0.7]{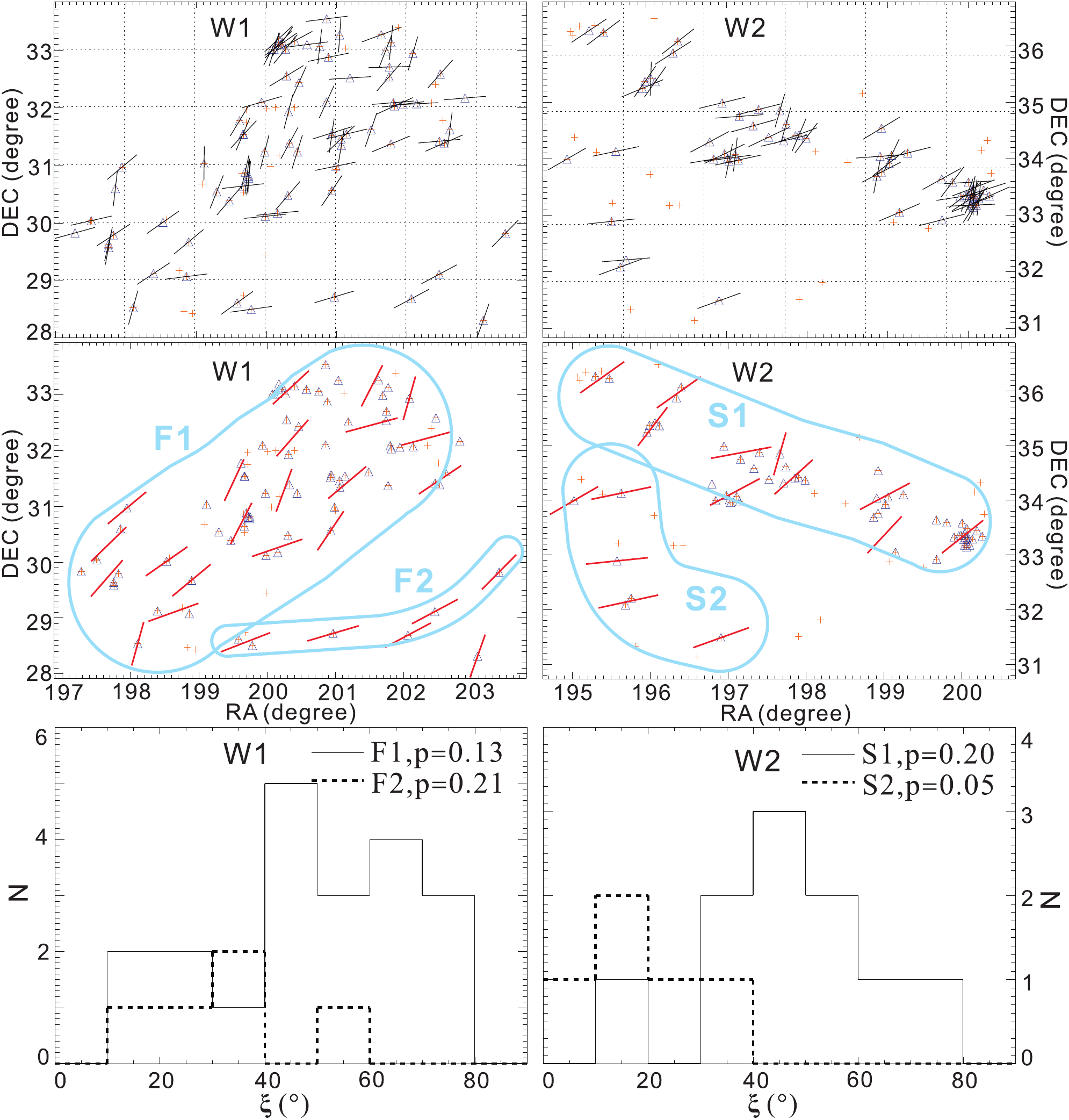}
\caption{The left and right panels show the orientations of the red galaxies and distribution of each filament in W1 and W2, respectively. The top panels show the orientations of the selected red galaxies with the $3\sigma$, which are denoted by the black bars. The dashed grids divide W1 and W2 into $1^{\circ}\times 1^{\circ}$ cells. The middle panels show the average orientations (denoted by the red bars) of the black bars in each cell. The four filaments F1, F2, S1, and S2 are circled by the blue lines. The bottom panels show the distribution histograms of the orientations of red bars in each filament region, and the results of K-S test are illustrated by the $p$ values.}
\label{f_ori}
\end{figure*}

Finally, we plot the orientations (denoted by the black bars) of the selected red galaxies by LAM with the $3\sigma$ confidence level for each wedge in the top panels of Fig.~\ref{f_ori}. In order to clearly show how many filaments have been detected and the overall orientations of the detected filaments, we divide the two wedges into $1^{\circ}\times1^{\circ}$ cells, as shown in Fig.~\ref{f_ori}, and calculate the average orientation of the selected red galaxies in each cell, which is denoted by a red bar in the middle panels of Fig.~\ref{f_ori}. The overall orientations of filaments are clearly revealed by the red bars. According to the average orientations, there are two main filaments in W1 (F1 and F2), and also two filaments, or more precisely, two sheets (S1 and S2; the filaments are located in the planes of the sheets; Falco et al. 2014; Tempel \& Libeskind 2013) in W2. These filaments are circled by the blue lines in Fig.~\ref{f_ori}. For each filament, we also plot the distribution of orientations of the red bars in the filament region, as shown in the bottom panels of Fig.~\ref{f_ori}. Kolmogorov-Smirnov test (K-S test) is performed to detect the deviation of each orientation distribution from a uniform distribution, and the $p$ values of all of the four distributions are small ($p=0.13$ for F1, $p=0.21$ for F2, $p=0.20$ for S1, and $p=0.05$ for S2), suggesting that the orientations of the red galaxies in each filament are indeed anisotropic.

\section{Discussion}

The method of \cite{Falco14} can only identify the filaments gravitationally bound to the clusters of galaxies, i.e., only the galaxies whose radial velocities are influenced by the gravity from the cluster matter can be detected; whereas LAM can theoretically find all of the filaments either influenced or not by the gravity of the clusters, since the orientations of red galaxies are not related directly to the gravity of the clusters of galaxies. Therefore the low value of $N_{\rm{c}}/N_{\rm{det}}$ in W1 might indicate that there are some galaxies located in filaments but not or only weakly influenced by the gravity of the Coma cluster in W1.

\subsection{Effect of alignment of red galaxies}

We are interested in whether the selected galaxies are due to both the alignments and inhomogeneous distributions of the galaxies in the 2D image, or just due to the latter. For example, perhaps the detected filament (sheet) S1 in W2 is identified because of the high number density of galaxies in the $360\--540\ \rm{arcmin}$ radius range rather than the alignments of red galaxies. In order to test the effect of alignment, the galaxies in the selected wedges themselves are treated as the background galaxies; meanwhile we artificially set $\xi$ of the background galaxies uniformly random in $0\--90^{\circ}$ and test the excess $m_i'=(n_i-n_i^{\rm{bg}})/n_i^{\rm{bg}}$ in each cell for $10^5$ times. The results are also listed in Table~\ref{N_comp}. We find that the filaments can also be identified with the $1\sigma$ confidence level; however the detection efficiency ($N_{\rm{c}}/N_{\rm{std}}$) is much worse than that by LAM. The effect of alignments can be characterized by the fraction $f=N_{\rm{c}}'/N_{\rm{c}}$, where $N_{\rm{c}}$ and $N_{\rm{c}}'$ are the numbers of detected galaxies by LAM and alignments only; $f=73.2\%$ for W1, and $f=66.7\%$ for W2. Therefore the galaxies selected by alignments account for substantial parts of the galaxies selected by LAM, suggesting that the alignments of red galaxies play an important role in LAM.

\subsection{LAM without redshift information}

The galaxies alignments and 2D distributions of galaxies are independent of the redshifts, therefore we expect that LAM should be still effective without the information of redshifts of the galaxies. However in this case, some interlopers, i.e., the foreground and background galaxies, will be included. The orientations of the foreground and background galaxies should be isotropic, if there is no strong gravitational lens in front of these galaxies, and the galaxies are not arranged in other filaments. Therefore the interlopers may be contaminations for LAM. With these in mind, we retrieve the catalog of galaxies around the Coma cluster (within $9^{\circ}$) from SDSS DR12 again, without restricting the range of redshifts. In order to reduce the contamination of the interlopers in the image, the range of the $r$ band magnitudes of the selected red galaxies need to be carefully chosen. Here we only use the bright red galaxies within $12\ {\rm{mag}}<m_r<16\ \rm{mag}$ to detect the filaments, where the faint end of $16\ \rm{mag}$ is brighter than the previous value of $18\ \rm{mag}$, since the alignments of the brighter red galaxies are more significant \citep{Tempel15}, and a brighter faint end induces fewer background interlopers; the other selecting criteria of the red galaxies remain unchanged. We apply LAM to the new sample of red galaxies, and the result for each wedge (W1 and W2) is also compared with the standard filaments, which is listed in the last column of Table~\ref{N_comp}. We find that the results for both W1 and W2 are slightly worse than the results of redshifts restricted; however, the detection efficiency (83.0\% for W1, 87.9\% for W2) is also acceptable. Additionally, $N_{\rm{c}}/N_{\rm{det}}$ (56.3\% for W1, 84.0\% for W2) for each wedge is also close to that with the redshifts restricted (58.3\% for W1, 79.7\% for W2), implying that the selected filaments without the redshifts information of galaxies are almost the same as those with the redshifts restricted.

We find that, in W1 and W2, 24 and 15 galaxies of the selected red galaxies by LAM (with $1\sigma$ confidence level) are not found without the redshifts information, respectively. Among these red galaxies which are not found without using the redshifts information, 18 (for W1) and 11 (for W2) galaxies are fainter than $m_r=16\ \rm{mag}$, suggesting that these galaxies are not found is mainly due to the fact that their magnitudes are out of range. Therefore the results without the redshifts information are slightly worse than the results of LAM, because some faint ($m_r>16\ \rm{mag}$) galaxies are not included in the sample ($12\ {\rm{mag}}<m_r<16\ \rm{mag}$) we used here.

\subsection{Alignment of ``background" galaxies?}

\begin{figure*}
\centering
\includegraphics[scale=0.7]{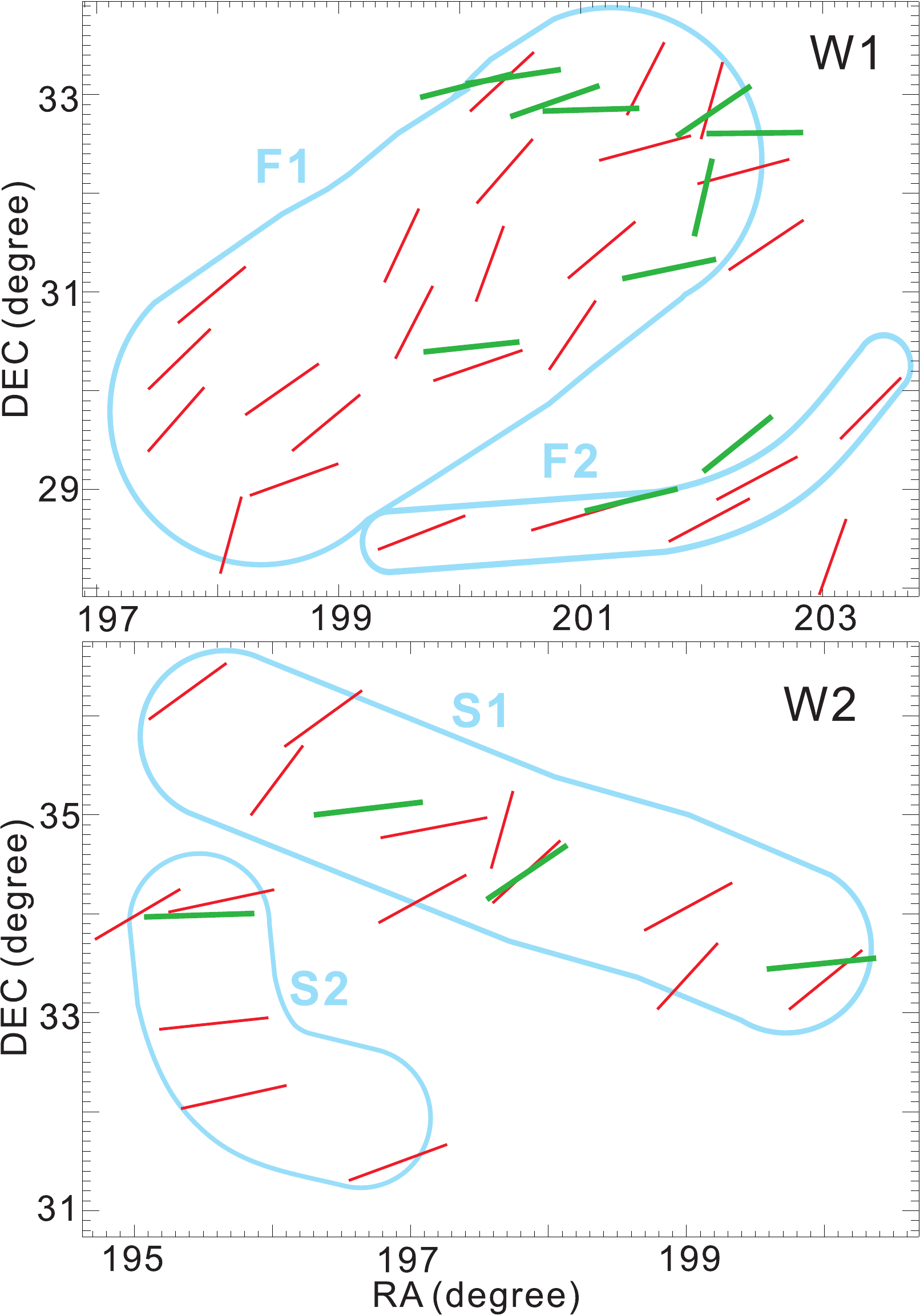}
\caption{The upper and lower panels show the orientations of ``background" interlopers in W1 and W2, respectively. The red bars denoting the overall orientations of filaments are the same as those in the middle panels of Fig.~\ref{f_ori}, and the green bars denote the orientations of the ``background" interlopers in the sample of selected red galaxies without restricting redshifts.}
\label{ori_inter}
\end{figure*}

Without restricting the redshifts of galaxies, we select $N_{\rm{det}}=87$ red galaxies in W1, and $N_{\rm{det}}=69$ galaxies in W2. Among these selected red galaxies, 72 (for W1) and 64 (for W2) red galaxies were found by LAM before, and only 15 (for W1) and 5 (for W2) red galaxies were not found by LAM before; meanwhile among these red galaxies which were not found by LAM, 11 (for W1) and 4 (for W2) galaxies are located at the redshifts $z>0.037$ (i.e., the given lower redshift limit of background galaxies), suggesting that almost all of the galaxies which were not found by LAM are possible background interlopers. We directly plot the orientations of the possible background interlopers for W1 and W2 in Fig.~\ref{ori_inter}, and find that the orientations of the possible background interlopers tend to be parallel to the overall orientations of filaments. In order to test whether the tendency is caused by physical reasons or just by accident, we need to study the orientations of the all possible background red galaxies ($z>0.037$, $12\ {\rm{mag}}<m_r<16\ \rm{mag}$) behind the four filaments/sheets.

\begin{figure*}
\centering
\includegraphics[scale=0.7]{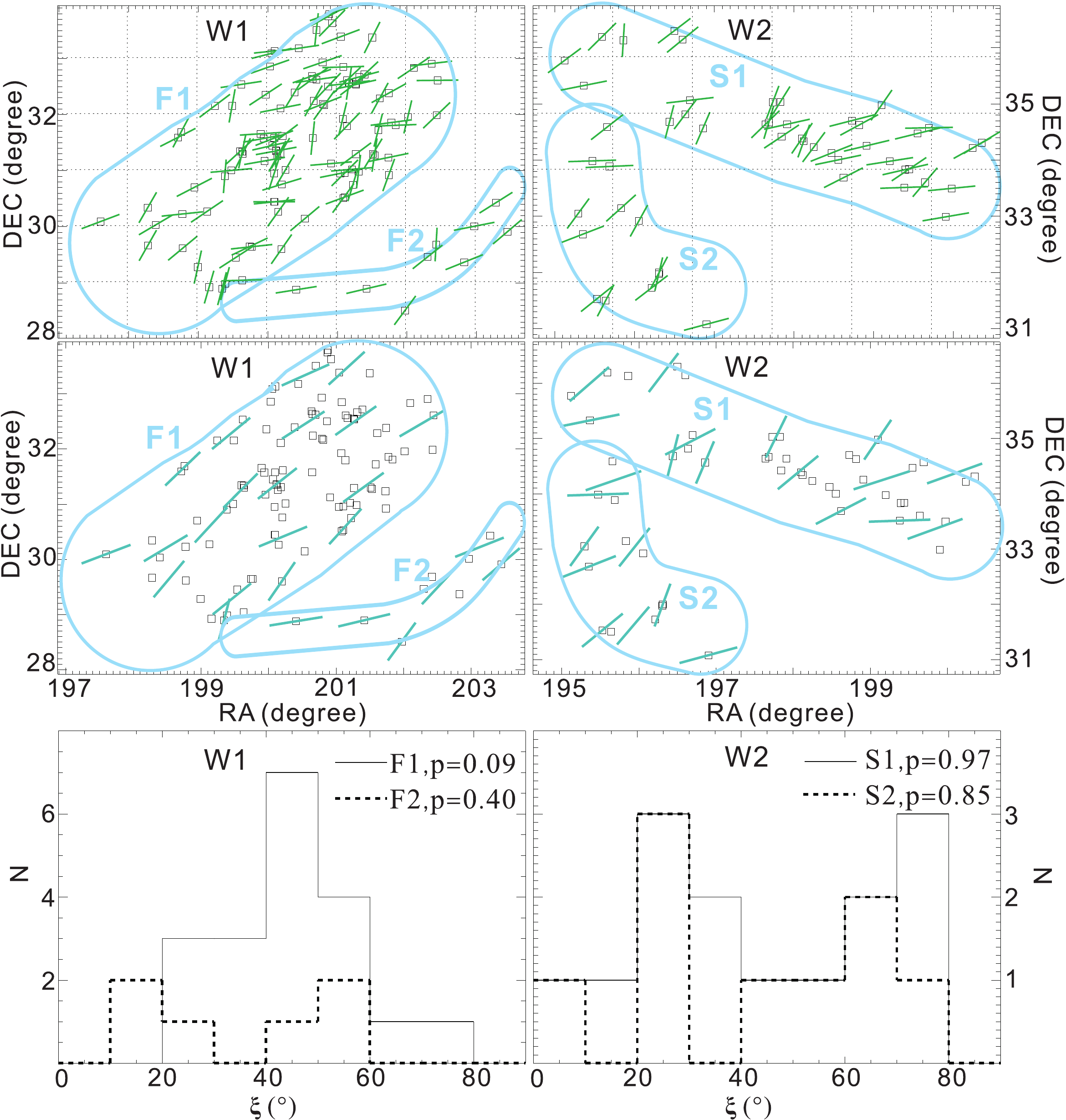}
\caption{The orientations of the red ``background" galaxies in each filament region and corresponding distribution of average orientations of the ``background" galaxies. Analogous to Fig.~\ref{f_ori}, the left and right panels correspond to W1 and W2, respectively. The top panels show the orientations of the red ``background" ($z>0.037$) galaxies in each filament region, which are denoted by the green bars. The dashed grids divide W1 and W2 into $1^{\circ}\times 1^{\circ}$ cells. The middle panels show the average orientations (denoted by the cyan bars) of the green bars in each cell. The four filament regions F1, F2, S1, and S2 are circled by the blue lines. The bottom panels show the distribution histograms of the orientations of cyan bars in each filament region, and the results of K-S test are illustrated by the $p$ values.}
\label{ori_bg}
\end{figure*}

First, we choose all possible background red galaxies ($z>0.037$, $12\ {\rm{mag}}<m_r<16\ \rm{mag}$) in the four filament regions in the 2D image as circled by the blue lines in Fig.~\ref{f_ori}, and plot their orientations in the upper panels of Fig.~\ref{ori_bg}. Second, we calculate the average orientation of the possible background galaxies in each $1^{\circ}\times1^{\circ}$ spatial cell, with the same manner as used for the filament galaxies (see section 2.3 for detail), and plot the average orientations in the middle panels of Fig.~\ref{ori_bg}. Finally, analogous to the lower panels in Fig.~\ref{f_ori}, we plot the distribution histogram of the average orientations in each filament region, and use K-S test to test the deviation from a uniform distribution; $p$ values are shown in the lower panels of Fig.~\ref{ori_bg}.

We find that the orientation distribution of the possible background galaxies in F1 region significantly deviates from a uniform distribution ($p=0.09$), whereas the distributions of possible background galaxies in F2, S1, and S2 regions are more likely to be uniform; meanwhile, the most significant excess in the average orientation distribution of the possible background galaxies behind F1 is located at about $40^{\circ}\--50^{\circ}$, which is consistent with the excess location of the red galaxies in F1 filament (see the lower left panel of Fig.~\ref{f_ori}). Therefore the selection of the possible background interlopers in F1 region by LAM is more likely due to a physical reason rather than accident. For F2 region, the uniform distribution ($p=0.40$) may be due to the small number of statistics. Therefore for detecting F2, the effect of the possible background galaxies is not clear. For S1 and S2 regions, the possible background interlopers are more likely to be selected by accident, because of the high values of $p$ ($p=0.97$ for S1 region, $p=0.85$ for S2 region). Therefore for detection of S1 and S2, the possible background interlopers are contaminations.

There are some possible mechanisms that may result in the relatively significant alignment signal of the possible background galaxies behind F1. For example, according to many previous papers about the effect of gravitational lensing, theoretically the images of background galaxies might be stretched along the orientation of the foreground filament by the shear of the filament (e.g., Higuchi et al. 2014). However, the gravitational lensing by filaments is extremely weak \citep{Dolag06,Mead10}, and changes the observed ellipticity of a background galaxy by only $2|\gamma|\simeq 0.01$ \citep{Waerbeke01}, where $\gamma$ is the shear of gravitational lensing. Therefore it is very unlikely that gravitational lensing contributes to the detected significant alignment signal shown in the lower left panel of Fig.~\ref{ori_bg}.

\begin{figure}
\centering
\includegraphics[scale=0.49]{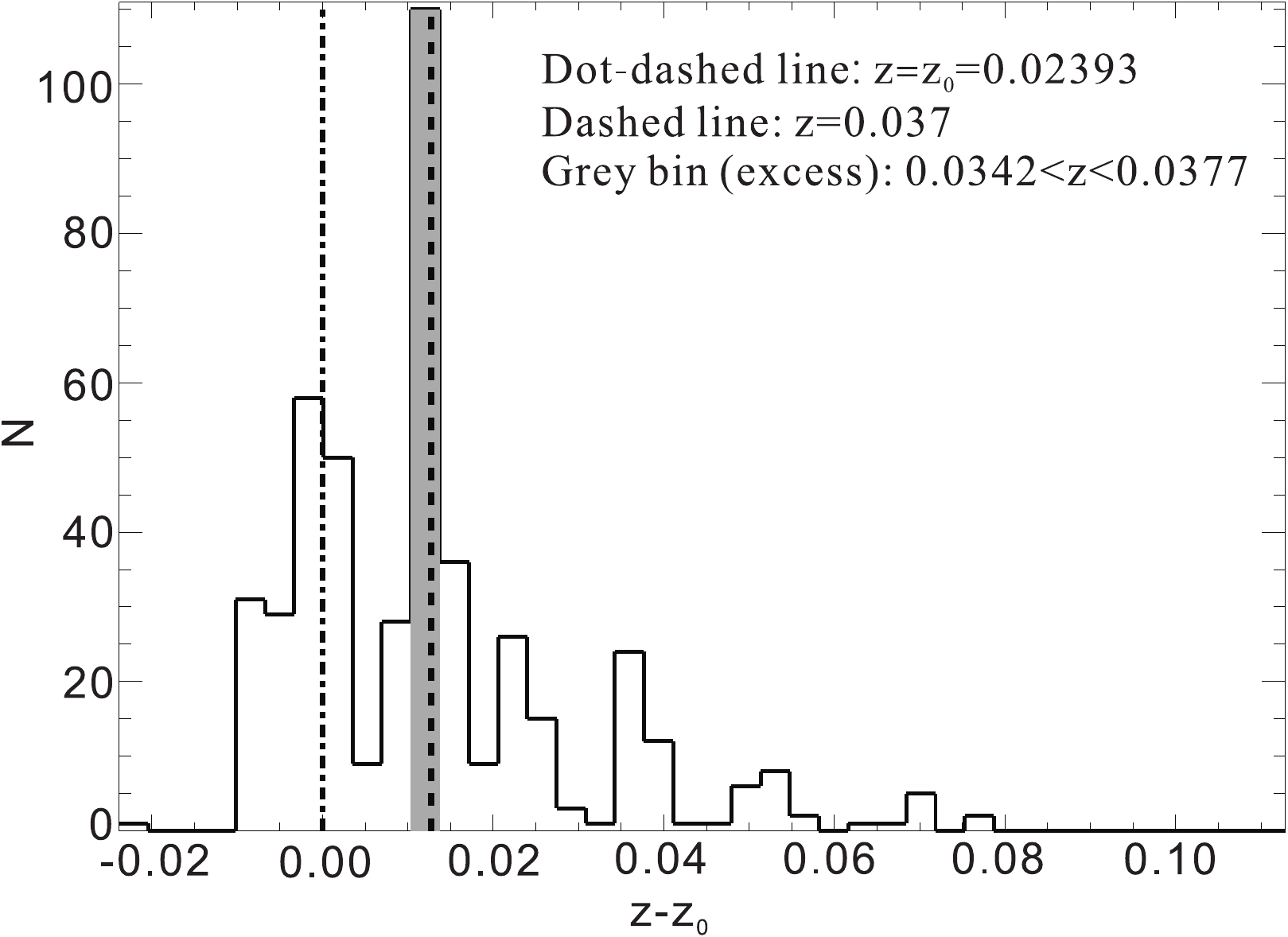}
\caption{Distribution histogram of relative redshifts of the all galaxies in F1 region in the 2D image. The dot-dashed and dashed lines denote $z=z_0=0.02393$ and $z=0.037$, respectively. The excess with the range of $0.0342<z<0.0377$ is denoted by the grey bin.}
\label{dis_z}
\end{figure}

Another possibility is that if some member galaxies in F1 are classified as background galaxies even if $z>0.037$, these galaxies may result in the detected alignment signal. In order to test this possibility, we plot the distribution of relative redshifts $z-z_0$ of all red galaxies in F1 region, as shown in Fig.~\ref{dis_z}; $z$ and $z_0=0.02393$ are the observed redshifts of a red galaxy and NGC~4874 (center of Coma cluster), respectively. The dot-dashed and dashed lines denote $z=z_0$ (i.e., the redshift of Coma center) and $z=0.037$ (i.e., the redshift lower limit of the possible background galaxies), respectively. If there is no filament in F1 region, we should expect a decreasing number of galaxies with increasing $z-z_0$, in the range of $z>z_0$. However, a significant excess is located at about $0.0342<z<0.0377$ (denoted by the grey bin in Fig.~\ref{dis_z}), indicating that there may be a filament (F1). Therefore a significant fraction of the member galaxies in F1 are indeed classified as ``background'' galaxies which may explain the significant alignment signal of ``background'' galaxies.

\begin{figure}
\centering
\includegraphics[scale=0.49]{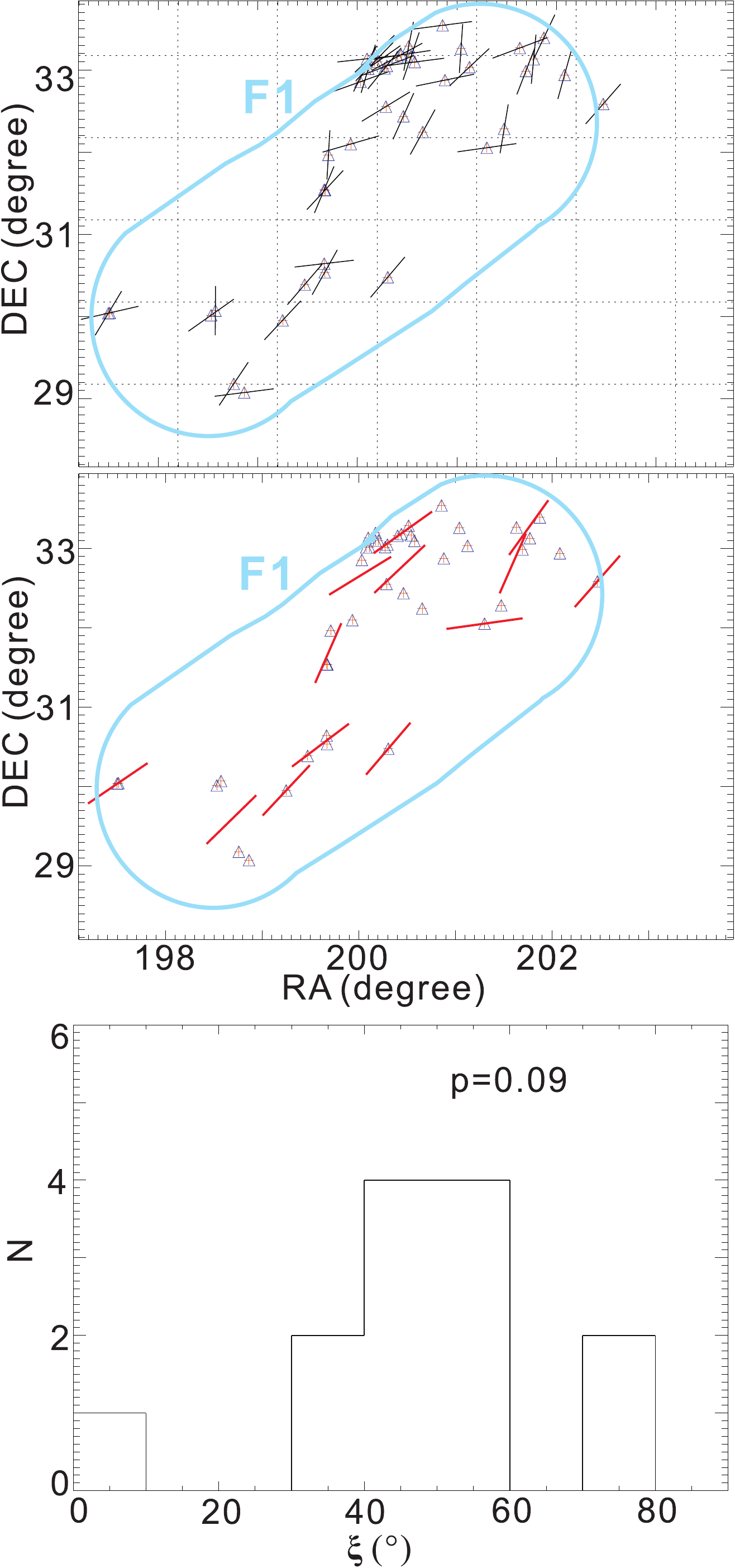}
\caption{Analogous to Fig.~\ref{f_ori}, from the upper to lower panels, we plot the orientations of the selected (with $3\sigma$ confidence) red galaxies with $0.0342<z<0.0377$ in F1 region, average orientation in each spatial cell, and distribution of the average orientations. The result of K-S test is illustrated by the $p$ value.}
\label{ori_nf}
\end{figure}

To further test whether the excess in Fig.~\ref{dis_z} really reveals a filament (F1), we use LAM again with the red galaxies in F1 region with $0.0342<z<0.0377$ (the redshift range of the excess in Fig.~\ref{dis_z}), and plot the orientations of the final selected red galaxies (with $3\sigma$ confidence) and distribution of orientations in Fig.~\ref{ori_nf}. We find that F1 is indeed detected with these galaxies, suggesting that the excess in Fig.~\ref{dis_z} is resulted from the member galaxies in F1. The $p$ value ($p=0.09$) is smaller than that with $0.01<z<0.037$ ($p=0.13$ as shown in Fig.~\ref{f_ori}), also indicating that the red galaxies with $0.037<z<0.0377$ in F1 region indeed are member galaxies of F1.

Therefore for W1, we need to subtract the galaxies with $0.037<z<0.0377$ from the eleven selected background galaxies shown in Fig.~\ref{ori_inter}. After removing these galaxies, we find that the contaminations by the left background interlopers are not strong when LAM is used without the information of galaxy redshifts, since there are only seven ($8.0\%$ of the total selected galaxies) and four ($5.8\%$ of the total selected galaxies) background interlopers in W1 and W2, respectively. Therefore, an advantage of LAM is that the method is independent of the redshifts of galaxies, and thus can be applied to detecting filaments at relatively high redshifts, where there are photometric images but lack of spectroscopic data and thus the previous algorithms of detecting filaments (e.g., using the velocities field) fail.

\section{Summary and Conclusion}

Since the red galaxies are preferentially aligned with their host filaments, we developed a new method, called location-alignment-method (LAM), of detecting filaments around clusters of galaxies, which uses both the alignments of the red galaxies and their distributions in 2D images. For the first time, the orientations of red galaxies are used to identify filaments. We applied LAM to the environment of Coma cluster, and compared our results with the ``standard'' filaments detected by the line-of-sight velocities and 2D positions of the galaxies \citep{Falco14}. We summarize our results as follows.

1) LAM can effectively find out the filaments around a cluster with $1\sigma$ confidence level, and even relatively higher confidence levels.

2) Four filaments (two filaments are located in sheets) are found in the two selected regions (W1 and W2),

3) The alignment of red galaxies is important in LAM, since a substantial part ($73.2\%$ for W1, and $66.7\%$ for W2) of the selected red galaxies is due to the alignment.

4) Applying LAM to the samples of bright (brighter than $16\ \rm{mag}$) red galaxies without the information of redshifts, we find that the filaments can still be detected (for W1 and W2, 83.0\% and 87.9\% red galaxies in the filaments are detected, respectively). The contaminatons by background interlopers are not strong.

In conclusion, there are two main advantages of LAM: (1) LAM can clearly reveal the number and overall orientations of the detected filaments; (2) LAM is independent of the redshifts of galaxies; therefore, we can use LAM to select the red galaxies in filaments among the sample of galaxies without the information of redshifts. Thus LAM can be applied at relatively high redshifts where there are photometric images but lack of spectroscopic data.

\section*{Acknowledgments}

We much thank the referee for his/her helpful discussion. SNZ acknowledges partial funding support by 973 Program of China under grant 2014CB845802, by the National Natural Science Foundation of China under grant Nos. 11133002 and 11373036, by the Qianren start-up grant 292012312D1117210, and by the Strategic Priority Research Program ``The Emergence of Cosmological Structures'' of the Chinese Academy of Sciences under grant No. XDB09000000.

\bibliographystyle{mn2e}


\end{document}